\documentclass[english,12pt]{article}
\usepackage{amssymb}
\usepackage{amsmath}
\usepackage{babel}
\usepackage[cp1250]{inputenc}
\date{}
\title{Quantum Game Theory in Finance}
\author{Edward W. Piotrowski\\ Institute of Theoretical Physics,
University of Bia\l ystok,\\ Lipowa 41, Pl 15424 Bia\l ystok,
Poland\\ e-mail: ep@alpha.uwb.edu.pl \\
 Jan S\l adkowski\\ Institute of Physics, University of Silesia, \\ Uniwersytecka
4, Pl 40007 Katowice, Poland \\ e-mail: sladk@us.edu.pl }
\begin{document}
\maketitle
\begin{abstract}
\noindent This is a short review of the background and recent
development in quantum game theory and its possible application in
economics and finance. The intersection of science and society is
also discussed. The review is addressed to non--specialists.
 \end{abstract}
 \vspace{5mm}

\section{Introduction}

One hundred years ago, a single concept  changed our view of the
world forever: quantum theory was born \cite{Zel}. Contemporary
technology is based on implementation of  quantum phenomena as a
result of this seminal idea. Regardless of the successes of
quantum physics and the resulting quantum technology social
sciences persist in classical paradigm what in some aspects can be
considered as an obstacle to unification of science in the quantum
domain. Quantum theory is up to now the only scientific theory
that requires the observer to take into consideration the usually
neglected influence of the method of observation on the result of
observation. Full and absolutely objective information about the
investigated phenomenon is impossible and this is a fundamental
principle of Nature and does not result from deficiency in our
technology or knowledge. Now, this situation is being changed in a
dramatic way. Fascinating results of quantum cryptography, that
preceded public key cryptography \cite{Wie} although not duly
appreciated at its infancy, caused that quantum information
processing is currently expanding its domain. Various proposals of
applying quantum--like models in social sciences and economics has
been put forward \cite{Baa}-\cite{nova}.  It seems that the
numerous acquainted with quantum theory physicists who have
recently moved to finance can cause an evolutionary change in the
paradigm of methods of mathematical finance. In a quantum world we
can explore plenty of parallel simultaneous evolutions of the
system and a clever final measurement may bring into existence
astonishing and classically inaccessible solutions \cite{nova}-\cite{NC}.
The price we are to pay consists in securing perfect
discretion to parallel evolution: any attempt (intended or not) at
tracing the system inevitably destroys the desirable quantum
effects. Therefore we cannot expect that all quantum aspects can
be translated and explained in classical terms \cite{hol} (if such
a reinterpretation was possible the balance could be easily
redressed). Attention to the very physical aspects of information
processing revealed new perspectives of computation, cryptography
and  communication methods.  In most of the cases quantum
description of the system provides advantages over the classical
situation. One should be not surprised that game theory, the study
of (rational) decision making in conflict situations, has quantum
counterpart. Indeed, games against nature \cite{mil} include those
for which nature is quantum mechanical. Does quantum theory offer
more subtle ways of playing games? Game theory considers
strategies that are probabilistic mixtures of pure strategies. Why
cannot they be intertwined in a more complicated way, for example
interfered or entangled? Are there situations in which quantum
theory can enlarge the set of possible strategies? Can quantum
strategies be more successful than classical ones? All these
questions have positive and sometimes bewildering answers
\cite{inv, nova}. There are genuine quantum games, that is games
that can be defined and played only in a sophisticated quantum
environment. Some of these quantum games could  be played only in
physical laboratories but technological development can soon
change this situation (the most interesting examples emerge from
cryptography). Some classical games can be redefined so that
quantum strategies can be adopted \cite{Mey}-\cite{IT4}. This is
ominous because someone can take the advantage of new (quantum)
technology if we are not on alert \cite{Mey, nova}. We should warn
the reader that quantum games are games in the classical sense but
to play a quantum game may involve sophisticated technology and
therefore  theoretical analysis of the game requires knowledge of
physical theories and phenomena necessary for its implementation.
This fact is often overlooked and quantum game theory is wrongly
put in sort of opposition to (classical) game theory. Recently, in
a series of papers \cite{PS1, PS2, PS6} the present authors
described market phenomena in terms of quantum game theory. Agents
adopting quantum strategies can make profits that are beyond the
range of classical markets. Quantum approach shed new light on
well known paradoxes \cite{PSN,Sla} and computational complexity
of economics \cite{vel, PSA}. Besides the properties of Nature
discovered by human beings there is a whole universe of phenomena
and appliances created by mankind. Therefore the question  if
present day markets reveal any (observable) quantum properties,
although interesting, is secondary to our main problem of finding
out if genuine quantum markets would ever come into existence.
Quantum theory offers a new paradigm that is able to produce a
unified description of reality. This paper is organized as
follows. First, we present some basic ideas of quantum games. Then
we describe quantum market games and review their attractive
properties. Finally we present our personal view of the further
development and possible applications of this field of research.

\section{Quantum market games}
As we have said in the Introduction, quantum game theory
investigates  conflict situations involving quantum phenomena.
Therefore it exploits the formalism of quantum theory. In this
formalism strategies are vectors (called states) in some Hilbert
space and can be interpreted as superpositions of trading
decisions. Tactics and moves are performed by unitary
transformations on vectors in the Hilbert space (states). The idea
behind using quantum games is to explore the possibility of
forming linear combination of amplitudes that are complex Hilbert
space vectors  (interference, entanglement \cite{nova}) whose
squared absolute values give probabilities of players actions. It
is generally assumed that a physical observable (e.g\mbox{.}
energy, position), defined by the prescription for its
measurement, is represented by a linear Hermitian operator. Any
measurement of an observable produces with some probability an
eigenvalue of the operator representing the observable. This
probability is given by the squared modulus of the coordinate
corresponding to this eigenvalue in the spectral decomposition of
the state vector describing the system. This is often an advantage
over classical probabilistic description where one always deals
directly with probabilities. The formalism has potential
applications outside physical laboratories \cite{Baa}-\cite{PS1}.
But how to describe complex games with say unlimited number of
players or non--constant pay--offs. There are several possible ways
of accomplishing this task. We have proposed a generalization of
market games to the quantum domain in Ref.~\cite{PS1}. In our
approach spontaneous or institutionalized market transactions are
described in terms of projective operation acting on Hilbert
spaces of strategies of the traders. Quantum entanglement is
necessary (non--trivial linear combinations of vectors--strategies
have to be formed) to strike the balance of trade. This approach
predicts the property of undividity of attention of traders (no
cloning theorem) and unifies  the English auction with the
Vickrey's one attenuating the motivation properties of the later
\cite{Sta}. Quantum strategies create unique opportunities for
making profits during intervals shorter than the characteristic
thresholds for an effective market (Brownian motion) \cite{Sta}.
Although the effective market hypothesis assumes immediate price
reaction to new information concerning the market the information
flow rate is limited by physical laws such us the constancy of the
speed of light. Entanglement of states allows to apply quantum
protocols of super--dense coding \cite{NC} and get ahead of
"classical trader". Besides, quantum version of the famous Zeno
effect \cite{NC} controls the process of reaching the equilibrium
state by the market.  Quantum arbitrage based on such phenomena
seems to be feasible. Interception of profitable quantum
strategies is forbidden by the impossibility of cloning of quantum
states. There are apparent analogies with quantum thermodynamics
that allow to interpret market equilibrium as a state with
vanishing financial risk flow. Euphoria, panic or herd instinct
often cause violent changes of market prices. Such phenomena can
be described by non--commutative quantum mechanics. A simple
tactics that maximize the trader's profit on an effective market
follows from the model: {\it accept profits equal or greater than
the one you have formerly achieved on average}\/ \cite{PS3}. \\

We were led to these conclusions by consideration of the following
facts:
\begin{itemize}
\item error theory: second moments of a random
variable describe errors,
\item H.~Markowitz's portfolio theory,
\item L.~Bachelier's theory of options:  the random variable $q^{2} + p^{2}$ measures joint risk
for a stock buying--selling transaction ( and Merton \& Scholes
works that gave them Nobel Prize in 1997).
\end{itemize}
We have defined canonically conjugate Hermitian operators
(observables) of demand $\mathcal{Q}_k$ and supply $\mathcal{P}_k$
corresponding to the variables $q$ and $p$ characterizing strategy
of  the $k$-th player. These operators act on the player's strategy
states $|\psi\rangle$\footnote{We use the standard Dirac notation.
The symbol $|\ \rangle$ with  a letter $\psi$ in it denoting a
vector parameterized by $\psi$ is called a {\it ket}; the symbol
$\langle\ |\negthinspace$ with a letter in it is called a {\it
bra}. Actually a {\it bra} is a dual vector to the corresponding
{\it ket}. Therefore scalar products of vectors take the form
$\langle \phi |\psi\rangle\negthinspace$ ({\it bracket}) and the
expectation value of an operator $A$ in the state
$|\psi\rangle\negthinspace$ is given by $\langle \psi
|A\psi\rangle\negthinspace$. A common abuse of this convention
consist in denoting  the wave function $\psi (p)$ as $\langle p
|\psi\rangle\negthinspace$. (A wave functions is a vector in
Hilbert space of square integrable functions and one associates
with the variable p
 an eigenvector $|p\rangle\negthinspace$.)} that have two important representations
$\langle q|\psi\rangle\negthinspace$ (demand representation) and
$\langle p|\psi\rangle\negthinspace$ (supply representation) where
$q$ and $p$ are logarithms of prices at which the player is buying
or selling, respectively \cite{NC,QAP}. This led us to the
following definition of the observable that we call {\it the risk
inclination operator} \cite{QAP}\footnote{The reader that is
familiar with the rudiments of quantum mechanics would certainly
notice that this operator is nothing else then the hamiltonian for
quantum harmonic oscillator.}: $$
H(\mathcal{P}_k,\mathcal{Q}_k):=\frac{(\mathcal{P}_k-p_{k0})^2}{2\,m}+
                     \frac{m\,\omega^2(\mathcal{Q}_k-q_{k0})^2}{2}\,,
\label{hamiltonian} $$ \noindent where
$p_{k0}\negthinspace:=\negthinspace\frac{
\phantom{}_k\negthinspace\langle\psi|\mathcal{P}_k|\psi\rangle_k }
{\phantom{}_k\negthinspace\langle\psi|\psi\rangle_k}\,$,
$q_{k0}\negthinspace:=\negthinspace\frac{
\phantom{}_k\negthinspace\langle\psi|\mathcal{Q}_k|\psi\rangle_k }
{\phantom{}_k\negthinspace\langle\psi|\psi\rangle_k}\,$,
$\omega\negthinspace:=\negthinspace\frac{2\pi}{\theta}\,$.  $
\theta$ denotes the characteristic time of transaction
\cite{PS3,QAP} which is, roughly speaking, an average time spread
between two opposite moves of a player (e.~g.~buying and selling
the same commodity). The parameter $m\negthinspace>\negthinspace0$
measures the risk asymmetry between buying and selling positions.
Analogies with quantum harmonic oscillator allow for the following
characterization of quantum market games. One can introduce an
analogue of the Planck constant, $h_E$, that describes the minimal
inclination of the player to risk, $
[\mathcal{P}_k,\mathcal{Q}_k]=\frac{i}{2\pi}h_E$. As the lowest
eigenvalue of the positive definite operator $H$ is
$\frac{1}{2}\frac{h_E}{2\pi} \omega$, $h_E$ is equal to the
product of the lowest eigenvalue of
$H(\mathcal{P}_k,\mathcal{Q}_k) $ and $2\theta$. $2\theta $ is in
fact the minimal interval during which it makes sense to measure
the profit. In a general case the operators $\mathcal{Q}_k $ do
not commute because traders observe moves of other players and
often act accordingly. One big bid can influence the market at
least in a limited time spread. Therefore it is natural to apply
the formalism of noncommutative quantum mechanics where one
considers $$ [ x^{i},x^{k}] = i \Theta ^{ik}:=i\Theta \,\epsilon
^{ik}.\eqno(3) $$ The analysis of harmonic oscillator in more than
one dimension \cite{qosc} imply that the parameter $\Theta $
modifies the constant $\hslash_E$ $\rightarrow \sqrt{\hslash_E^{2}
+ \Theta ^{2}} $ and the eigenvalues of
$H(\mathcal{P}_k,\mathcal{Q}_k)$ accordingly. This has the natural
interpretation that moves performed by other players can diminish
or increase one's inclination to take risk. Encouraged by that we
asked the question {\it Provided that an all--purpose quantum
computer is built, how would a market cleared by a quantum
computer perform?} To find out we have to consider quantum games
with unlimited and changing number of players. A possible approach
is as follows. If a game allows a great number of players in it is
useful to consider it as a two--players game: the $k$-th trader
against the Rest of the World (RW). Any concrete algorithm
$\mathcal{A}$ should allow for an effective strategy of RW (for a
sufficiently large number of players the single player strategy
should not influence the form of the RW strategy). Let the real
variable $q$ $$q:= \ln c - E(\ln c) $$ denotes the logarithm of
the price at which the $k$-th player can buy the asset
$\mathfrak{G}$ shifted so that its expectation value in the state
$\mid \psi > _{k}$ vanishes. The expectation value of $x$ is
denoted by $E(x)$. The variable $p$ $$p:= E(\ln c) - \ln c  $$
describes the situation of a player who is supplying the asset
$\mathfrak{G}$ according to his strategy $|\psi\rangle_k$.
Supplying $\mathfrak{G}$ can be regarded as demanding $\$$ at the
price $c^{-1}$ in the $1\mathfrak{G}$ units and both definitions
are equivalent. Note that we have defined $q$ and $p$ so that they
do not depend on possible choices of the units for $\mathfrak{G}$
and $\$ $. For simplicity we will use such units that $E(\ln c)
=0$. The strategies $|\psi \rangle_{k}$ belong to  Hilbert spaces
$H_{k}$. The state of the game $|\Psi\rangle_{in}:=\sum_k|\psi
\rangle_k$ is a vector in the direct sum of Hilbert spaces of all
players, $\oplus _{k} H_k$. We will define canonically conjugate
hermitian operators of demand $\mathcal{Q}_k$ and supply
$\mathcal{P}_k$ for each Hilbert space $H_{k}$ analogously to
their physical position and momentum counterparts. This can be
justified in the following way. Let $\exp(-p)$ be a definite
price, where $p$ is a proper value of the operator
$\mathcal{P}_k$. Therefore, if one have already declared the will
of selling exactly at the price $\exp(-p)$ (the strategy given by
the proper state $|p\rangle_{k}$) then it is pointless to put
forward any opposite offer for the same transaction. The capital
flows resulting from an ensemble of simultaneous transactions
correspond to the physical process of measurement. A transaction
consists in a transition from the state of traders strategies
$|\Psi\rangle_{in}$ to the one describing the capital flow state
$|\Psi\rangle_{out}:=\mathcal{T}_\sigma |\Psi\rangle_ {in}$, where
$\mathcal{T}_{\sigma}:=\sum_{k_d}|q\rangle_{k_d}\phantom{}_{k_d}
   \negthinspace\langle q|+
 \sum_{k_s}|p\rangle_{k_s}\phantom{}_{k_s}
   \negthinspace\langle p|$  is the projective operator defined by
the division $\sigma $ of the set of traders $\{ k\}$ into two
separate subsets $\{k\}=\{k_d\}\cup\{k_s\}$, the ones buying at
the price $e^{q_{k_d}}$ and the ones selling at the price
$e^{-p_{k_s}}$ in the round of the transaction in question. The
role of the algorithm $\mathcal{A}$ is to determine the division
$\sigma$ of the market, the set of price parameters $\{ q_{k_{d}},
p_{k_{s}}\}$ and the values of capital flows. The later are
settled by the distribution $$\int_{-\infty}^{\ln c}
\frac{{|\langle q|\psi\rangle_k|}^2}{\phantom{}_k
\negthinspace\langle\psi|\psi\rangle_k}dq $$ which is interpreted
as the probability that the trader $| \psi \rangle _{k}$ is
willing to buy  the asset $\mathfrak{G}$ at the transaction price
$c$ or lower \cite{PS3}. In an analogous way the distribution $$
\int_{-\infty}^{\ln \frac{1}{c}} \frac{{|\langle
p|\psi\rangle_k|}^2}{\phantom{}_k
\negthinspace\langle\psi|\psi\rangle_k}dp  $$ gives the
probability of selling $\mathfrak{G}$ by the trader $| \psi
\rangle _{k}$ at the price $c$ or greater. These probabilities are
in fact conditional because they describe the situation after the
division $ \sigma $ is completed. If one considers the RW strategy
it make sense to declare its simultaneous demand and supply states
because for one player RW is a buyer and for another it is a
seller. To describe such situation it is convenient to use the
Wigner formalism \footnote{Actually, this approach consists in
allowing pseudo--probabilities into consideration. From the
physical point of view this is questionable but for our aims its
useful, c.f\mbox{.} the discussion of the Giffen paradox \cite{Sla}.}
\cite{wigner}. The pseudo--probability $W(p,q)dpdq$ on the phase
space $\{(p,q)\}$ known as the Wigner function is given by
\begin{eqnarray*}
W(p,q)&:=& h^{-1}_E\int_{-\infty}^{\infty}e^{i\hslash_E^{-1}p x}
\;\frac{\langle
q+\frac{x}{2}|\psi\rangle\langle\psi|q-\frac{x}{2}\rangle}
{\langle\psi|\psi\rangle}\; dx\\
&=& h^{-2}_E\int_{-\infty}^{\infty}e^{i\hslash_E^{-1}q x}\;
\frac{\langle
p+\frac{x}{2}|\psi\rangle\langle\psi|p-\frac{x}{2}\rangle}
{\langle\psi|\psi\rangle}\; dx,
\end{eqnarray*}
where the positive constant $h_E\negthinspace=\negthinspace2\pi\hslash_E$ is the
dimensionless economical counterpart of the Planck constant.
Recall that this measure is not positive definite except for the
cases presented below. In the general case the pseudo--probability
density of RW is a countable linear combination of Wigner
functions, $\rho(p,q)=\sum_n w_n W_n (p,q)$,
 $w_n\geq 0$, $\sum_n w_n =1$.
 The diagrams of the integrals of the RW
pseudo--probabilities (see Ref\mbox{.} \cite{PS3}) $$ F_d(\ln
c):=\int_{-\infty}^{\ln c} \rho(p={const.},q)dq $$ (RW bids
selling at $\exp {(-p)}$)\\and $$ F_s(\ln c):=\int_{-\infty}^{\ln
\frac{1}{c}} \rho(p,q={const.})dp $$ (RW bids buying at $\exp(q)$)
against the argument $\ln c$ may be interpreted as the dominant
supply and demand curves in the Cournot convention, respectively
\cite{PS3}. Note, that due to the lack of positive definiteness of
$\rho $, $F_d$ and $F_s$ may not be monotonic functions. Textbooks
on economics give examples of such departures from the law of
supply (work supply) and law of demand (Giffen assets) \cite{sam}.
We proposed to call an arbitrage algorithm resulting in non
positive definite probability densities {\it a giffen}. The
following subsection describe shortly various aspects of quantum
markets.

\subsection{Quantum Zeno effect} It has been experimentally verified that sufficiently
frequent measurement can slow down (accelerate) the dynamics of a
quantum proces, what is called the {\it quantum (anti--)Zeno
effect} \cite{Zenon}. Analogous phenomenon can be observed in
quantum games. If the market continuously measures the same
strategy of the player, say the demand $\langle q|\psi\rangle $,
and the process is repeated sufficiently often for the whole
market, then the prices given by the algorithm $\mathcal{A}$ do
not result from the supplying strategy $\langle p|\psi\rangle $ of
the player. The necessary condition for determining the profit of
the game is the transition of the player to the state $\langle
p|\psi\rangle $ \cite{PS3}. If, simultaneously, many of the
players change their strategies then the quotation process may
collapse due to the lack of opposite moves. In this way the
quantum Zeno effects explain stock exchange crashes. Effects of
this crashes should be predictable because the amplitudes of the
strategies $\langle p|\psi\rangle$  are Fourier transforms of
$\langle q|\psi\rangle$. Another example of the quantum market
Zeno effect is the stabilization of prices of an asset provided by
a monopolist.

\subsection{ Eigenstates of $\mathcal{Q}$ and $\mathcal{P}$}

Let us suppose that the amplitudes for the strategies $\langle
q|\psi\rangle_{k}$ or $\langle p|\psi\rangle _{k}$  have divergent
integrals of their modulus squared. Such states live outside the
Hilbert space but have the natural interpretation as the desperate
will of the $k$-th player of buying (selling) of the amount $ d_k$
( $s_k$) of the asset $\mathfrak{G}$. So the strategy $\langle
q|\psi\rangle_k= \langle q|a\rangle =\delta(q,a)$ means, in the
case of classifying the player into the set $\{k_d\}$, refusal of
buying cheaper than at $c=e^{a}$ and the will of buying at any
price equal or higher than $e^{a}$. In the case of a "measurement"
in the set $\{k_d\}$ the player declares the will of selling at
any price. The above interpretation is consistent with the
Heisenberg uncertainty relation. The strategies $\langle
q|\psi\rangle_2=\langle q|a\rangle$ (or $\langle
p|\psi\rangle_2=\langle p|a\rangle$) do not correspond to the RW
behaviour because the conditions $d_2,s_2>0$, if always satisfied,
allow for unlimited profits (the readiness to buy or sell
$\mathfrak{G}$ at any price). The appropriate demand  and supply
functions give probabilities of coming off transactions in a game
when the player use the strategy $\langle p|{const}\rangle$ or
$\langle q|{const}\rangle$ and RW, proposing the price, use the
strategy $\rho$ \cite{PS1,PS3}. The authors have analyzed the
efficiency of the strategy $\langle q|\psi\rangle_1=\langle
q|-a\rangle$ in a two--player game when RW use the strategy with
squared modulus of the amplitude equal to normal distribution
\cite{PS3}. The maximal intensity of the profit \cite{PS3} is
equal to 0.27603 times the variance of the RW distribution
function. Of course, the strategy $\langle p|\psi\rangle_1=\langle
p|0,27603\rangle$ has the same properties. In such  games
a=0.27603 is a global fixed point of the profit intensity
function. This may explain the universality of markets on which a
single client facing the bid makes up his/hers mind. Does it mean
that such common phenomena have quantal nature? The Gaussian
strategy of RW \cite{temp} can be parameterized by a
temperature--like parameter $T=\beta^{-1}$. Any decrease in
profits is only possible by reducing the variance of RW
(i.e\mbox{.}  cooling). Market competition is the mechanism
responsible for the risk flow that allows the market to attain the
"thermodynamical"  balance. A warmer market influences
destructively the cooler traders and they diminish the uncertainty
of market prices.

\subsection{Correlated coherent strategies}

We will define correlated coherent strategies as the eigenvectors
of the annihilation operator $\mathcal{C}_k$ \cite{dod} $$
\mathcal{C}_k(r,\eta):=\frac{1}{2\eta}\Bigl(1+\frac{ir}{\sqrt{1-r^2}}
\Bigr)\mathcal{Q}_k + i\eta\mathcal{P}_k , $$ where $r$ is the
correlation coefficient  $r\in[-1,1]$, $\eta>0$. In these
strategies buying and selling transactions are correlated and the
product of dispersions fulfills  the Heisenberg--like uncertainty
relation $\Delta_p\Delta_q\sqrt{1-r^2}\geq\frac{\hslash_E}{2}$ and
is minimal. The annihilation operators $\mathcal{C}_k$ and their
eigenvectors may be parameterized by
$\Delta_p=\frac{\hslash_E}{2\eta}$,
$\Delta_q=\frac{\eta}{\sqrt{1-r^2}}$, and $r$. This leads to
following form of the correlated Wigner coherent strategy $$
W(p,q)dpdq=\frac{1}{2\pi\Delta_p\Delta_q\sqrt{1-r^2}}\;e^{-\frac{1}{2(1-r^2)}
\bigl(\frac{(p-p_0)^2}{\Delta^2_p}+\frac{2r(p-p_0)(q-q_0)}{\Delta_p\Delta_q}+
\frac{(q-q_0)^2}{\Delta^2_q}\bigr)}dpdq\,. $$ They are not giffens.
It can be shown, following Hudson \cite{Hud}, that they form the
set of all pure strategies with positive definite Wigner
functions. Therefore pure strategies that are not giffens are
represented in phase space $\{(p,q)\}$ by gaussian distributions.

\subsection{ Mixed states and thermal strategies}

According to classics of game theory \cite{NM} the biggest choice
of strategies is provided by the mixed states $\rho(p,q)$. Among
them the most interesting are the thermal ones. They are
characterized by constant inclination to risk,
$E(H(\mathcal{P},\mathcal{Q}))={const}$ and maximal entropy. The
Wigner measure for the $n$-th exited state of harmonic oscillator
has the form \cite{Tatar} $$
W_n(p,q)dpdq=\frac{(-1)^n}{\pi\hslash_E}\thinspace
e^{-\frac{2H(p,q)}{\hslash_E\omega}}
L_n\bigl(\frac{4H(p,q)}{\hslash_E\omega}\bigr)dpdq, $$ where
$L_{n}$ is the $n$-th Laguerre polynomial. The mixed state
$\rho_\beta$ determined by the Wigner measures $W_ndpdq$ weighted
by the Gibbs distribution $w_n(\beta):=\frac{e^{-\beta
n\hslash_E\omega}}{\sum_{k=0}^\infty e^{-\beta k\hslash_E\omega}}$
has the form
\begin{eqnarray*}
\rho_\beta (p,q)dpdq:&=&\sum_{n=0}^\infty w_n(\beta) W_n(p,q)
dpdq\\ &=&\frac{\omega}{2\pi}\;x\; e^{-xH(p,q)}
\Bigr|_{x\frac{2}{\hslash_E\omega}\tanh(\beta\frac{\hslash_E\omega}{2})}
dpdq .\end{eqnarray*}
So it is a two dimensional normal
distribution. It easy to observe that by recalling that
$\frac{1}{1-t}\thinspace e^\frac{xt}{t-1}=\sum_{n=0}^\infty
L_n(x)t^n$ is the generating function for the Laguerre
polynomials. It seems to us that the above distributions should
determine the shape of the supply and demand curves for
equilibrium markets. There are no giffens on such markets. It
would be interesting to investigate the temperatures of
equilibrium markets. In contrast to the traders temperatures
\cite{temp} which are Legendre coefficients and measure "trader's
qualities" market temperatures are related to risk and are
positive. The Feynman path integrals may be applied to the
Hamiltonian  to obtain equilibrium quantum Bachelier model of
diffusion of the logarithm of prices of shares that can be
completed by the Black--Scholes formula for pricing European
options \cite{hul}.
\subsection{Quantum auctions and bargaining}
After tasting the exotic flavour  of quantum market games one may
wish to distinguish the class of quantum transactions
(q-transactions) that is q-games without institutionalized
clearinghouses. This class includes quantum bargaining
(q-bargaining)  and quantum auctions (q-auction). The participants
of a q-bargaining game will be called Alice ($A$) and Bob ($B$).
We will suppose that they settle on beforehand who is the buyer
(Alice) and who is the seller (Bob). A two--way q-bargaining that
is a q-bargaining when the last condition is not fulfilled can be
treated analogously. Alice enter into negotiations with Bob to
settle the price for the transaction. Therefore the proper
measuring apparatus consists of the pair of traders in question.
In q-auction the measuring apparatus consists of a one side only,
the initiator of the auction. We showed \cite{PS1} that the
players strategies can be described in terms of polarizations,
that is the states in a two--dimensional Hilbert space. If the
player formulates the conditions of the transaction we say she has
the polarization ${\mit 1}$ (and is in the state
$|\overrightarrow{r}\rangle_A=|{\mit 1}\rangle$). In q-bargaining
this means that she puts forward the price. In the opposite case,
when she decides if the transaction is made  or not, we say she
has the polarization $|{\mit 0}\rangle$\/. (She accepts or not the
conditions of the proposed transaction.) There is an analogy of
the isospin symmetry in nuclear physics which says that nucleon
has two polarization states: proton and neutron. The vectors
$(|{\mit 0}\rangle,|{\mit 1}\rangle)$\/ form an orthonormal basis
in $\mathcal{H}_{\it{s}}$\/, the linear hull of all possible Alice
polarization states.  The player $1$ proposes a price and the
player -1 accepts or reject the proposal. Therefore their
polarizations  are $|\mit0\rangle$ and $|\mit1\rangle$,
respectively so the q-bargaining has  the polarization
$|\mit0\rangle _{\text{-}1}|\mit1\rangle _{1}$\cite{PS1,PS2}. The
transaction in question is accomplished if the obvious rationality
condition is fulfilled
\begin{equation*}
\label{haucja-war} [{\mathfrak q}+{\mathfrak p}\leq 0]\,,
\end{equation*}
where  the convenient Iverson notation \cite{GKP} is used
($[expression]$\/ denotes the logical value (1 or 0) of the
sentence $expression$) and the parameters ${\mathfrak p}=\ln
\mathfrak{c}_{\text{-}1}$\/ and $-{\mathfrak q}=\ln {\mathfrak
c}_1$\/ are random variables corresponding to prices at which the
respective players withdraw, the {\sl withdrawal prices}. The
variables $\mathfrak{p}$ and $\mathfrak{q}$ describe (additive)
profits resulting from price variations. Their probability
densities are equal to squared absolute values of the appropriate
wave functions $\langle p|\psi\rangle_{\text{-}1}$\/ and $\langle
q|\psi\rangle_1$ (that is their strategies).  Note that the
discussed q-bargaining may result from a situation where
several players have intention of buying but they were outbid by
the player $1$ (his withdrawal price ${\mathfrak c}_1$\/ was
greater than the  other players ones,
 ${\mathfrak c}_1>{\mathfrak c}_k$,
$k=2,\ldots,N$)\/. This means that all part in the auction behave
like fermions (e.g\mbox{.} electrons) and they are subjected to a sort of
Pauli exclusion principle according to which two players cannot
occupy the same state. This surprising statement consist in
noticing that the transaction in question is made only if the
traders have opposite polarizations (and even that is not a
guarantee of the accomplishment). The fermionic character of
q-bargaining parts was first noted in \cite{PS1} in a slightly
different context. If at the outset of the auction there are
several bidding players then the rationality condition takes the
form
\begin{equation*}
\label{haucja-war2} [{\mathfrak q}_{\min} +{\mathfrak p}\leq 0]
\end{equation*}
where ${\mathfrak q}_{\min}
:=\underset{k=1,\ldots,N}{\min}\{{\mathfrak q}_k\}$ is the
logarithm of the highest bid multiplied by $-1$. According to
Ref\mbox{.} \cite{PS1} the probability density of making the
transaction with the $k$-th buyer at the price $c_{k}={\mathrm
e}^{-q_k}$\/ is given by
\begin{equation}
\label{haucja-dobicie}
dq_k \;\frac{|\langle q_k|\psi_k\rangle |^2}{
\langle\psi_k|\psi_k\rangle} \prod_{\substack{m=1\\ m\neq k}}^{N}
\int_{-\infty}^{\infty}\negthinspace\negthinspace\negthinspace
dq_m \frac{|\langle q_m|\psi_m\rangle |^2
}{\langle\psi_m|\psi_m\rangle}\int_{-\infty}^{\infty}\negthinspace\negthinspace\negthinspace
dp\; \frac{|\langle p|\psi_{\text{-}1}\rangle
|^2}{\langle\psi_{\text{-}1}|\psi_{\text{-}1}\rangle}
\;[\;q_k=\negthinspace\negthinspace\min_{n=1,\ldots,N}\{q_n\}\;]\;[q_k+p\leq0]\,.
\end{equation}
The seller is not interested in making the deal with any
particular buyer and the unconditional probability of
accomplishing the transaction at the price $c$ is given by the sum
over $k\negthinspace=\negthinspace1,\ldots,N$\/ of the above
formula with $q_k\negthinspace=\negthinspace-\ln c$. If we neglect
the problem of determining the probability amplitudes in
$(\ref{haucja-dobicie})$\/ we easily note that the discussed
q-bargaining is in fact an English auction (first price
auction), so popular on markets of rare goods. It is interesting
to note that the formula $(\ref{haucja-dobicie})$\ contains wave
functions of payers who were outbid before the end of the
bargaining (cf the Pauli exclusion principle). The probability
density of "measuring" of a concrete value $q$ of the random
variable $\mathfrak{q}$ characterizing the player, according to
the probabilistic interpretation of quantum theory, is equal to
the squared absolute value of the normalized wave function
describing his strategy
\begin{equation*}
\label{cica-eisert} \frac{|\langle q|\psi_k\rangle|^2}{
\langle\psi_k|\psi_k\rangle}\,dq\;.
\end{equation*}
Physicists normalize wave functions because conservation laws
require that. Therefore the trivial statement that if a market
player may be persuaded into striking a deal or not is a matter of
price alone, corresponds to the physical fact that a particle
cannot vanish without any trace. The analysis of an English
q-auction with reversed roles that is with selling bidders is
analogous. The case when the polarization of the q-auction is
changed to $|\mit1\rangle_{\text{-}1}|\mit0\rangle_1$ is more
interesting. In this case the player -1 reveals her withdrawal
price and the player 2 accepts it (as the rest of the players do)
or not. Such an auction is known as the Vickrey's auction (or the
second price auction). The winner is obliged to pay the second in
decreasing order price from all the bids (and the withdrawal price
of the player -1). In the quantum approach English and Vickrey's
auctions are only special cases of a phenomenon called q-auction.
In the general case both squared amplitudes
$|\langle{\mit0_{\text{-}1}1_{1}}|{\mit0_{\text{-}1}1_{1}}\rangle|^2$
and
$|\langle{\mit1_{\text{-}1}0_{1}}|{\mit1_{\text{-}1}0_{1}}\rangle|^2$
are non--vanishing so we have to consider them with weights
corresponding to these probabilities. Such a general q-auction
does not have  counterparts on the real markets. It should be very
interesting to analyse the motivation properties of q-auctions eg
finding out when the best strategy is the one corresponding to the
player's valuation of the good.
If we consider only positive definite probability measures then
the bidder gets the highest profits in Vickrey's auction using
strategies with public admission of his valuation of the auctioned
good. But it might not be so for giffen strategies because
positiveness of measures is supposed in proving the incentive
character of Vickrey's auctions \cite{Klem}.  The presence of
giffens on real markets might not be so abstract as it seems to
be. Captain Robert Giffen who supposedly found additive measures
not being positive definite but present on  markets in the forties
of the XIX century \cite{Stig} probably got ahead of physicists in
observing quantum phenomena. Such departures from the demand law,
if correctly interpreted, does not cause any problem neither for
adepts nor for beginners. Employers have probably always thought
that work supply as function of payment is scarcely monotonous.
The distinguished by their polarization first and second price
auctions have analogues in the Knaster solution to the pragmatic
fair division problem (with compensatory payments for indivisible
parts of the property) \cite{Luc}. Such a duality might also be
found in election systems that as auctions often take the form of
procedures of solving fair division problems \cite{fair}. It might
happen that social frustrations caused by election systems would
encourage us to discuss such topics.

\section{Conclusions}

The commonly accepted universality of quantum theory should
encourage physicist in looking for traces of quantum world in
social phenomena. We envisage markets cleared by quantum computer.
We hope that the sketchy analysis presented above would allow the
reader to taste the exotic flavours  of quantum markets. A quantum
theory of markets provides new tools that can be used to explain
of the very involved phenomena including interference of (quantum)
strategies \cite{kecaJ} and diffusion of prices \cite{dyf}. The
research  into the quantum nature of games may offer solutions to
very intriguing paradoxes present in philosophy and economics. For
example, the Newcomb's paradox analyzed in Ref\mbox{.} \cite{PSN}
suggests various ones. There are quantum games that live across
the border of our present knowledge. For example, consider some
classical or quantum problem $X$. Let us define the game $kXcl$:
you win if and only if you solve the problem (perform the task)
$X$ given access to only $k$ bits of information. The quantum
counterpart reads: solve the problem $X$ on a quantum computer or
other quantum device given access to only $k$ bits of information.
Let us call the game $kXcl$ or $kXq$ interesting if the
corresponding limited information--tasks are feasible. Let
$OckhamXcl$ ($OckhamXq$) denotes the minimal $k$ interesting game
in the class $kXcl$ ($kXq$). Authors of the paper \cite{PS5}
described the game played by a market trader who gains the profit
$P$ for each bit (qubit) of information about her strategy. If we
denote this game by $MP$ then
$OckhamM\frac{1}{2}\,cl\negthinspace=\negthinspace2M\frac{1}{2}\,cl$
and for $P\negthinspace>\negthinspace\frac{1}{2}$ the game
$OckhamMPcl$ does not exist. They also considered the more
effective game $1M\frac{2+\sqrt{2}}{4}\,q$ for which
$OckhamM\frac{2+\sqrt{2}}{4}\,q\neq1M\frac{2+\sqrt{2}}{4}\,q$ if
the trader can operate on more then one market. This happens
because there are entangled strategies that are more profitable
\cite{EWP}. There are a lot of intriguing questions that can be
ask, for example for which $X$ the meta--game $Ockham(OckhamXq)cl$
can be solved or when, if at all,  the meta--problem
$Ockham(OckhamXq)q$ is well defined problem. Such problems arise
in quantum memory analysis \cite{Koe}.  We would like to stress
that this field of research undergoes an eventful development.
Therefore now it is difficult to predict which results would turn
out to be fruitful and which would have only marginal effect.

Recent research on the (quantum) physical aspects of information
processing should result in a sort of {\it total quantum paradigm}
and we dare to say that quantum game approach became sooner or
later a dominant one. Therefore we envisage markets cleared by
quantum algorithms (computers)\footnote{Let us quote the Editor's
Note to Complexity Digest 2001.27(4) (http://www.comdig.org):
 "It might be that while observing the
due ceremonial of everyday market transaction we are in fact
observing capital flows resulting from quantum games eluding
classical description. If human decisions can be traced to
microscopic quantum events one would expect that nature would have
taken advantage of quantum computation in evolving complex brains.
In that sense one could indeed say that quantum computers are
playing their market games according to quantum rules".}.
\\[1ex]{\bf Acknowledgements}: This paper has been supported by the
{\bf Polish Ministry of Scientific Research and Information
Technology} under the (solicited) grant No\mbox{.} {\bf
PBZ-MIN-008/P03/2003}.

\end{document}